# SPACECRAFT SWARM ATTITUDE CONTROL FOR SMALL BODY SURFACE OBSERVATION

Ravi teja Nallapu,[*] and Jekan Thangavelautham[†]

Understanding the physics of small bodies such as asteroids, comets, and planetary moons will help us understand the formation of the solar system, and also provide us with resources for a future space economy. Due to these reasons, missions to small bodies are actively being pursued. However, the surfaces of small bodies contain unpredictable and interesting features such as craters, dust, and granular matter, which need to be observed carefully before a lander mission is even considered. This presents the need for a surveillance spacecraft to observe the surface of small bodies where these features exist. While traditionally, the small body exploration has been performed by a large monolithic spacecraft, a group of small, low-cost spacecraft can enhance the observational value of the mission. Such a spacecraft swarm has the advantage of providing longer observation time and is also tolerant to single point failures. In order to optimize a spacecraft swarm mission design, we proposed the Integrated Design Engineering & Automation of Swarms (IDEAS) software which will serve as an end-to-end tool for theoretical swarm mission design. The current work will focus on developing the Automated Swarm Designer module of the IDEAS software by extending its capabilities for exploring surface features on small bodies while focusing on the attitude behaviors of the spacecraft in the swarm. We begin by classifying spacecraft swarms into 5 classes based on the level of coordination. In the current work, we design Class 2 swarms, whose spacecraft operate in a decentralized fashion but coordinate for communication. We demonstrate the Class 2 swarm in 2 different configurations, based on the roles of the participating spacecraft. The attitude behaviors of all the spacecraft are then converted to a line of sight (LoS) tracking problem with respect to different targets depending on their role in the swarm. A sliding mode control law is used to track the LoS with respect to assigned targets. Following this, we formulate the surface feature problem as an optimization problem which is solved using genetic algorithm optimization. Finally, the principles described are demonstrated by a numerical simulation of observing a simulated surface feature over the surface of asteroid 433 Eros. The results indicate successful performance of the design and control algorithms.

## INTRODUCTION

The hundreds of thousands small bodies in the solar system present exciting opportunities for exploration. Surface exploration of these small bodies has been discussed to answer high-value science questions[1] which are highlighted by the recent planetary science decadal survey[2]. Additionally, exploration of small bodies can provide insight into finding resources to sustain a future space economy[3,4,5,35,36]. Spacecraft architectures such as orbiters, landers and rovers are ideal for exploring small bodies as they have long duration access to a large portion of the target body. However, the design of such spacecraft faces few key challenges: Firstly, the physical characteristics of these bodies are poorly understood, and therefore orbital reconnaissance of these targets is needed to de-risk surface missions. Secondly, the dynamical motion of the spacecraft around the asteroid constrains the feasible orbits[6, 7]. Therefore, flyby observations are the ideal precursors for in-depth exploration and prospecting of these small bodies. Typically, flyby observations are carried out by a single spacecraft. However, a single spacecraft flyby can only provide coverage over a short duration. For this reason, we advocate flyby of multiple spacecraft or swarms for a detailed study of the target body. This work focuses on application of spacecraft swarms to study surface features on a target small body.

---

[*] PhD Candidate, Aerospace and Mechanical Engineering, University of Arizona.
[†] Assistant Professor, Aerospace and Mechanical Engineering, University of Arizona.

The design of any swarm spacecraft mission requires making design decisions and trade-offs at several hierarchical levels. This includes making (often) unintuitive decisions at the spacecraft level, at the level of the spacecraft swarm and the swarm trajectory level. Our approach is to develop an end-to-end design tool that automates the actual design process. Through automation, we may be able to obtain creative designs never thought of by the experimenter. To this effect, we propose the Integrated Design Engineering & Automation of Swarms (IDEAS) software[8]. The software will contain a knowledge base of swarm structure/organization, trajectory and spacecraft design and behavior, and will provide an optimal or near-optimal design using Evolutionary algorithms[9] (EA). The proposed software architecture of the IDEAS software is presented in Figure 1.

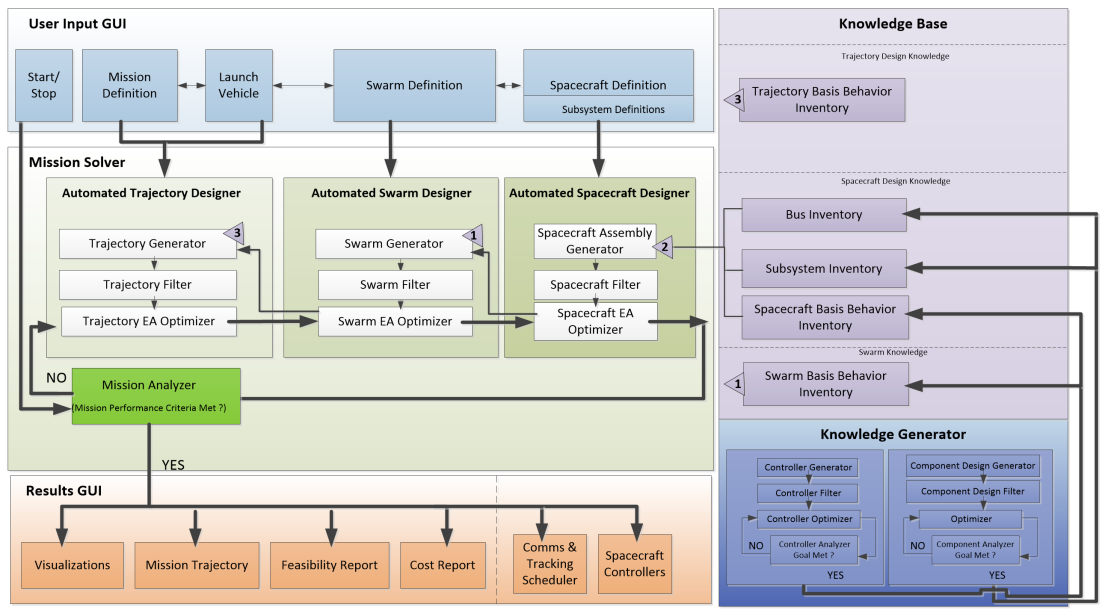

**Figure 1.** Software architecture of the IDEAS software which will serve as an end-to-end design tool for design spacecraft swarm missions to small bodies.

In the present work, we apply the Automated Swarm Designer module within IDEAS to design a swarm that provides detailed surveillance of a region of interest on a target body only using flybys. Specifically, a portion of the target body will be assumed to contain a surface feature of interest. The location of the feature and its spatial distribution is assumed to be known but a detailed surface map of the feature is required. We assume that spacecraft in the swarm have 2 functions: sensing and communication. The spacecraft which perform sensing will be responsible for observing the target feature with a camera and will be called the "Observing Spacecraft", while the spacecraft responsible for communication will relay the data back to a ground station on Earth and will be known as the "Communications Spacecraft". The 2 types of spacecraft are illustrated in Figure 2.

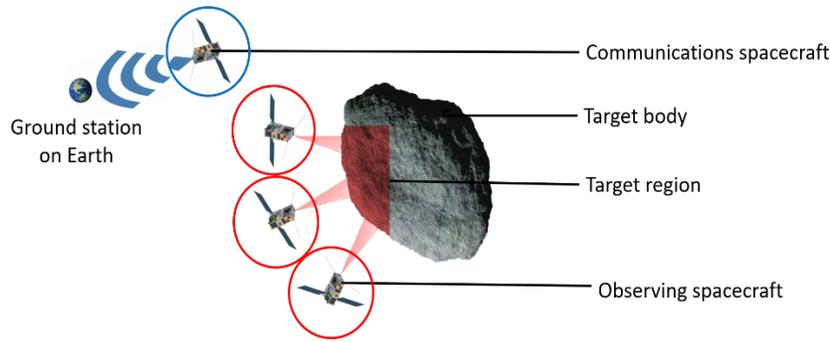

**Figure 2.** The two types of spacecraft populating the swarm include the "Observing Spacecraft" and the "Communication Spacecraft".

As the "Observing Spacecraft" will fly by the target body in the vicinity of the target region, they will be commanded to observe the target region along their Line of sight (LoS). Once the operation is complete, the spacecraft will communicate the data by pointing their antenna towards the "Communication Spacecraft" through low-frequency data transfer. This maneuver requires the "Observer Spacecraft" to orient their communication system towards the "Communications spacecraft." The "Communication Spacecraft" can then transfer this data back to an Earth-based ground station by maneuvering it antennas towards Earth. Furthermore, we study two configurations of such a swarm. The first configuration has a dedicated "Communications Spacecraft," while the second configuration has adaptable spacecraft that can determine their role based on a group selection scheme. The attitude behaviors of all the spacecraft will be tracking the Line of Sight (LoS) with respect to a moving target. A sliding mode control law will be used by the spacecraft to ensure the attitude tracking operation.

The organization of the paper is as follows. The next section will present related work on spacecraft swarms where we present a new classification for different types of swarms. Following this, we present the modeling methodologies used in the current work to design a spacecraft swarm for flyby mapping of a small-body. Then, numerical simulation is presented of a swarm mapping a location on asteroid 433 Eros. Finally, conclusions and future work is presented.

**RELATED WORK**

Spacecraft swarms have the potential to introduce whole new capabilities to space exploration and prospecting. Although there is no bound on the number of spacecraft required to constitute a swarm, we treat any collection of multiple spacecraft as a swarm. This section will present the relevant research done in the field of multi-spacecraft technology. Multi-spacecraft missions are broadly classified into 2 types: formation flying and constellations[10]. Formation flying missions aim to couple the dynamics of the constituent spacecraft so that they operate in some form of synchrony. Depending on the architecture, the formation flying missions are further classified into 2 types: centralized control[11], and decentralized control[12]. Centralized control architectures exhibit a central spacecraft, also known as the leader spacecraft, which computes the reference states and control laws of other participants in the swarm. While in the decentralized architectures, the constituent spacecraft make their own decisions. Constellations, on the other hand, require no coordination between their participants[13].

Modern research on spacecraft swarm guidance, navigation, and control (GNC) has focused on challenges such as the development of control laws for formation maintenance, robustness, cooperation, and swarm navigation. The formation maintenance problem is also known as the swarm keeping problem and has been well studied in the literature[14-18].

While formation flying has many practical applications, it is not a requirement for a swarm. Applications such as surface coverage/prospecting and persistent observations of target sites can be accomplished by an architecture that does not require each spacecraft maintain formation. Therefore, such applications can be designed through swarm constellations. Constellation design research has focused on maximization of payload spatial and temporal coverage[13, 19]. Currently, constellations are designed using the grid point method[13] where the target region is specified by a grid of points on the target body surface. The performance of constellations of different shapes and sizes are then tested either by varying them manually[20] or through optimization schemes[21]. However, constellation swarms have not been well studied for mapping small bodies. At the time of this work, spacecraft swarms have been considered as platforms to explore main belt asteroids utilizing distributed sensor networks[22, 23], mother-daughter configurations[24] and as gravimetry platforms for asteroids flybys[25, 26]. The dynamics of spacecraft around irregular bodies such as asteroids and comets are also being studied[27,28].

Currently, there is yet to be a unifying scheme for fast mapping of small bodies utilizing multiple spacecraft swarms. For us to have a unified treatment of swarms and lay the groundwork for distinguishing critical swarm behaviors, we introduce a new classification of spacecraft swarms based on the level of interaction between the participating spacecraft. We categorize a spacecraft swarm into 5 classes which are explained as follows:

*Class 0 Swarm*: It is simply a collection of multiple spacecraft that exhibit no coordination either in movement, sensing, or communication.

*Class 1 Swarm*: Each spacecraft coordinates its movement resulting in formation flying but there is no explicit communication coordination or sensing coordination.

*Class 2 Swarm*. Each spacecraft coordinates movement and communication including using Multiple-Input-Multiple-Output (MIMO) or parallel channels. The swarm has collective sensing capabilities but is not optimized with respect to the swarm layout or is post-processed.

*Class 3 Swarms*. Each spacecraft coordinates sensing/perception with communication and positioning/movement but is not collectively optimized. Individual losses can have uneven outcomes including total loss of the system.

*Class 4 Swarms*. Each spacecraft exploits concurrent coordination of positioning/movement, communication and sensing to perform system level optimization. The system acts if it's a single entity. Communication, computation and sensing is evenly distributed within the swarm. Individual losses result in gradual loss in system performance.

In our previous work, we focused on the design of Class 0 and Class 1 swarms[8, 20]. The present work will focus on designing Class 2 swarms where each spacecraft coordinate its movement and communication.

## METHODOLOGY

This section describes the methodology used in the current work to observe a region of interest on the target body. We present the surface observation problem and introduce its constraints. We then proceed to describe attitude behaviors of the 2 configurations of the Class 2 swarm being designed. We finally illustrate the Automated Swarm Design module within IDEAS software and show the design optimization problem that will be solved by an Evolutionary Algorithm.

**Surface feature observation**

In this current work, we assume that a target body contains a surface feature of interest whose surface spread is already known. The objective then is to observe the target region with a required ground resolution of $D$. For this reason, we will design a swarm with Observation and Communications Spacecraft, which flyby a target region for detailed surveillance. All the spacecraft in the swarm will be deployed from a carrier "mothership" spacecraft. The "Observation Spacecraft" will be responsible for observing the target region with the required ground resolution $D$ using a simulated camera as a payload. Once the spacecraft are past imaging distance, they transmit their data to the Communication Spacecraft. We will demonstrate the Class 2 swarm through two configurations of the swarm: In the first configuration, the swarm has a dedicated Communications Spacecraft that always maintains its LoS with respect to Earth. If $N_{Obs}$ denotes the number of Observing spacecraft in the swarm, then the total size of the swarm $N_{Sw} = N_{Obs} + 1$ in Configuration 1. In the second design, the swarm is assumed to be made of identical adaptable spacecraft which can play the role of an Observation and/or a Communications Spacecraft. The attitude behavior in this case will require that all spacecraft track the LoS with respect to the target body when inside the imaging region. Once outside the imaging region, the swarm can select the "Communications Spacecraft" based on a clustering scheme. In the current work, we assume that the 1$^{st}$ spacecraft in the swarm is the selected spacecraft that adapts to become the "Communications Spacecraft" outside the imaging region. Once the "Communications Spacecraft" is selected, the same behavior from the first configuration is followed. In this case, the swarm size is $N_{Sw} = N_{Obs}$.

**Attitude behaviors**

*Configuration 1.* Let $\bar{r}_j$ denote the position vector of the $j^{th}$ spacecraft with respect to the target body. The LoS vector of this spacecraft with respect to the target body is $\bar{A}_j = -\bar{r}_j$. The direction of this LoS vector is denoted by the unit vector $\hat{A}_j$. Similarly, the LoS vector of the spacecraft with respect to the "Communications Spacecraft" is denoted by $\hat{C}_j$. Let the body axis of this spacecraft be denoted by $\hat{z}_j$, and the imaging radius around the target body be denoted by $R_{im}$. Then the desired attitude behavior of the "Observation Spacecraft" can be described as follows: The body $\hat{z}_j$ axis tracks $\hat{A}_j$ if $|\bar{r}_j| \leq R_{im}$, otherwise the body $\hat{z}_j$ axis tracks $\hat{C}_j$. Similarly, if we denote the body $+z$ axis of the "Communications Spacecraft" as $\hat{z}_c$ and the LoS vector of the "Communications Spacecraft" to Earth as $\hat{E}$, then the required attitude behavior of the spacecraft is to make the $\hat{z}_c$ vector track $\hat{E}$.

*Configuration 2.* In this configuration, Spacecraft 1 orients its $\hat{z}_1$ axis along $\hat{A}_1$ if $|\bar{r}_1| \leq R_{im}$, and switches to track $\hat{E}$ outside the region. The remaining spacecraft align their $\hat{z}_j$ axis along $\hat{A}_j$ if $|\bar{r}_j| \leq R_{im}$. Outside the imaging region, they align their $\hat{z}_j$ axis along their $\hat{C}_j$ vector towards the spacecraft 1.

The behavior of the two configurations is illustrated in Figure 3. It can be seen here that the attitude behaviors of spacecraft require them to align their body frame $+z$ axis along the LoS with respect to a moving object. Representing the attitude by the Modified Rodrigues Parameters (MRP)[29] $\boldsymbol{\sigma}$, the current work uses a sliding mode control law[30] to track their reference LoS with respect to their assigned targets[31].

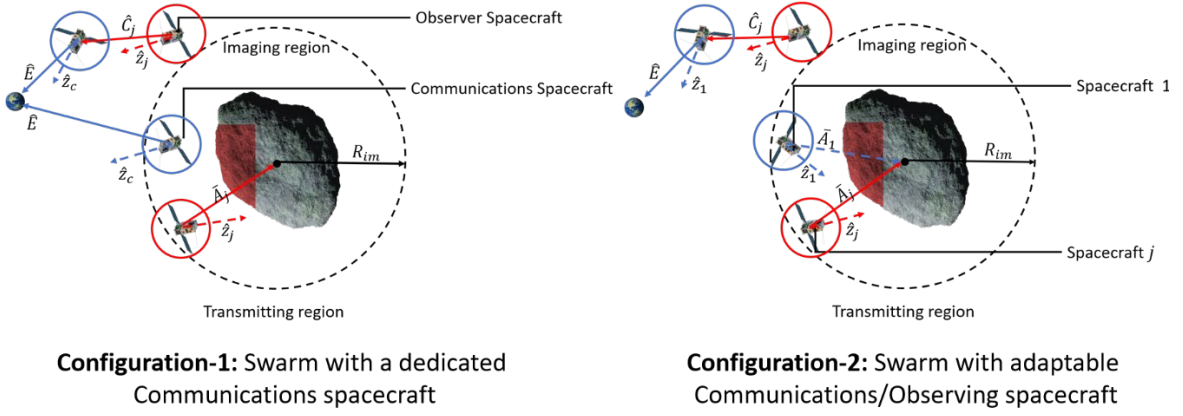

**Figure 3.** Attitude behaviors of the spacecraft in the swarm for the 2 configurations considered in the study.

**Swarm design**

*Target region generation.* The shape model of the target body is used to generate a virtual event. The shape model contains a vertex distribution of the target body $V_A$, and a face distribution array $F_A$, that lists the order of connections between the vertices required to construct the triangular faces of the target body. We generate the surface feature centered at a target location (latitude and longitude) by constructing a virtual pyramid that subtends a desired angle at the center of the target body. The culled and clipped faces that fall inside this virtual pyramid define the target region faces[8] $F_T \subset F_A$.

*Observation criterion.* The camera field of view (FoV) of an "Observation Spacecraft" is modeled as a pyramid originating at the spacecraft center, with its axis projected along the spacecraft $+z$ axis. The faces of the target body that fall within the FoV pyramid of the camera are selected based on a culling and clipping algorithm[8] and are used to define the observed face set $F_O \subset F_A$. The target area observed by the spacecraft swarm is then defined by the target set $F_{TO}$ as

$$F_{TO} = F_O \cap F_T \tag{1}$$

Finally, the area observed by a region, $A(F)$ is obtained by computing the sum of areas of the triangles described the face set. The figure of merit used in the current work to describe the target region observed by the swarm is given by the percentage area observed

$$P_{TO} = \frac{A(F_{TO})}{A(F_T)} \times 100 \;\% \tag{2}$$

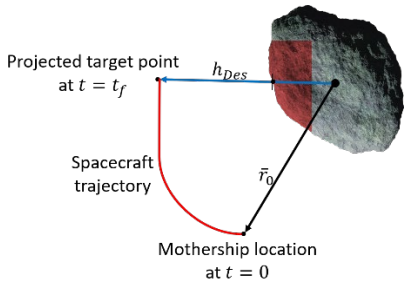

**Figure 4.** Illustration of the trajectory finding scheme.

*Trajectory generation.* The current work uses a trajectory design algorithm which requires the specification of the starting and final locations denoted by $\bar{r}_0$ and $\bar{r}_f$, and a time of flight $t_f$ for travel[8]. The starting point of all the spacecraft in the swarm is the location of the mothership. The destination points are the projections of the vertices involved in the definition of $F_T$, at a desired imaging altitude $h_{Des}$ above the surface of the target body. Let $n_T$ define the number of vertices described by $F_T$. Each spacecraft decides on one of the $n_T$ points through an optimization problem described by (3)-(6). In the first configuration, the communication spacecraft is commanded to visit the mean location of all the "Observation Spacecraft" at $t = t_f$. An illustration of the trajectory generation problem is shown in Figure 4.

*Optimization problem*. We will now cast the design of the target region observing swarm as an optimization problem. We are interested in minimizing the number of "Observation Spacecraft," $N_{Obs}$ and determining their location of visit which is specified by an index $p_j$ corresponding to the $j^{th}$ spacecraft. Therefore, the problem can be expressed as:

$$\min_{N_{Obs},\ p_j} N_{Obs} \quad (3)$$

such that

$$|100 - P_{TO}| - \epsilon \leq 0 \quad (4)$$
$$1 \leq p_j \leq n_T \quad \forall j \leq N_{Obs} \quad (5)$$

and

$$1 \leq N_{Obs} \leq N_{max} \quad \forall j \leq N_{Obs} \quad (6)$$

Equations (3)-(6) are solved using an Evolutionary Algorithm solver[32] in MATLAB. The design variable of the swarm is presented in a gene map format in Figure 5. A numerical simulation of the swarm design and its attitude behaviors is presented in the next section.

| Description | # Observing spacecraft | Spacecraft 1 Visiting point | Spacecraft 2 Visiting point | ... | Spacecraft $N_{Obs}$ Visiting point |
|---|---|---|---|---|---|
| Symbol | $N_{Obs}$ | $p_1$ | $p_2$ | | $p_{N_{Obs}}$ |
| Domain | Integers [1, $N_{max}$] | Integers [1, $n_T$] | | | |

**Figure 5.** Gene map of the designed swarm variable used in the present study.

## NUMERICAL SIMULATIONS

In this section, we demonstrate the automated swarm design process described in the previous section through numerical simulations. We will create a virtual surface feature on the surface of the asteroid 433 Eros, which the spacecraft are required to observe with a ground resolution of 10 cm. We design the swarm through the optimization problem in Equations (3)-(6) and present the attitude profiles of the participating spacecraft. The heliocentric orbits of 433 Eros and Earth are propagated using their fixed shape size parameters[33]. The gravitational environment around 433 Eros is modeled using a $2 \times 2$ spherical harmonic gravity field[27, 28].

### Surface feature

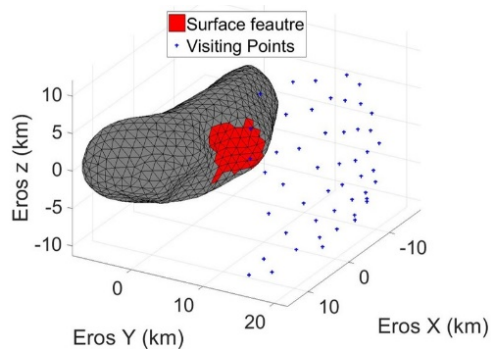

**Figure 6.** Generated surface event, and the projected visiting points on the target body

A surface feature of interest centered at 0 deg latitude, 90 deg on the surface of the asteroid, subtending a half angle of 20 deg angle is modeled. The surface feature corresponded to a total of 49 vertices. The required flyby altitude and required half cone angle of the camera was determined using coverage relations for an optical instrument[13] as 14 km and 21.5 deg respectively. The surface feature and the corresponding projected visiting points at the flyby altitude are presented in Figure 6. The mothership was assumed to be located at $[0\ \ 0\ \ -30]^T$ km in the rotating reference frame of the asteroid, at the start of the simulation. The time of flight from the mothership to a selected visiting point was specified as $t_f = 10$ mins. The operation was propagated for a total time of 20 mins to capture the attitude behaviors of all the spacecraft involved.

**Spacecraft design parameters**

All the spacecraft in the swarm are presumed to be 3U CubeSats[34] with a uniformly distributed mass of 8 kg. The spacecraft were initialized with random attitudes. The components of the MRP were distributed uniformly in $[-1, 1]$ while the components of angular velocities were distributed uniformly in $[-0.2, 0.2]$ RPM. The MRPs were propagated through a shadow set switching scheme to avoid singularities[29].

**Swarm design**

As mentioned earlier, the optimization problem in Equations (3)-(6) was solved using the Evolutionary Algorithms solver in MATLAB. The value of $N_{max} = 20$ spacecraft, and a tolerance of $\epsilon = 0.1\,\%$ was used to run the search problem. The algorithm ran for a total of 67 generations exploring a total of 6801 designs. The average and best fitness of each generation, which indicates the number of "Observation Spacecraft," is presented in Figure 7. As seen here, a minimum of $N_{Obs} = 5$ spacecraft can observe the target feature completely. The selected optimal swarm design is presented in the gene map format in Figure 8.

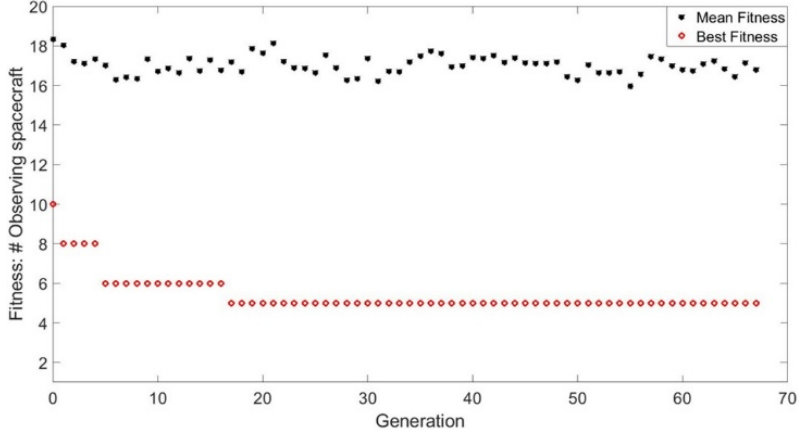

**Figure 7.** Results of the genetic algorithm optimization of the swarm design problem.

| # Observing spacecraft | Spacecraft 1 Visiting point | Spacecraft 2 Visiting point | Spacecraft 1 Visiting point | Spacecraft 2 Visiting point | Spacecraft 2 Visiting point |
|---|---|---|---|---|---|
| 5 | 7 | 25 | 31 | 34 | 38 |

**Figure 8.** The selected optimal design of the swarm expressed in the gene map format.

**Attitude control**

In this subsection, we present the tracking errors of the fittest solution. The tracking errors are expressed as the MRP difference[29] between the reference between the LoS reference and the actual attitude of the spacecraft,

*Configuration-1.* The MRP tracking errors of the spacecraft in Configuration 1 are presented in Figure 9. As seen here, the tracking errors are asymptotically shown to converge to 0. As expected the Observation Spacecraft enter into a feature observing mode as they are inside the imaging region. Once they enter the transmitting region, they switch their behavior and enter into a communication track mode (CT) where they track the LoS with respect to the Communications Spacecraft.

This switch in the attitude behavior results in an initial attitude spike which is shown to asymptotically converge to 0 due to the control torque. The Communications Spacecraft, on the other hand, is shown to be in the Earth tracking (ET) mode as the error MRP defining the alignment error between spacecraft $+z$ axis and the LoS with respect to Earth is smoothly regulated.

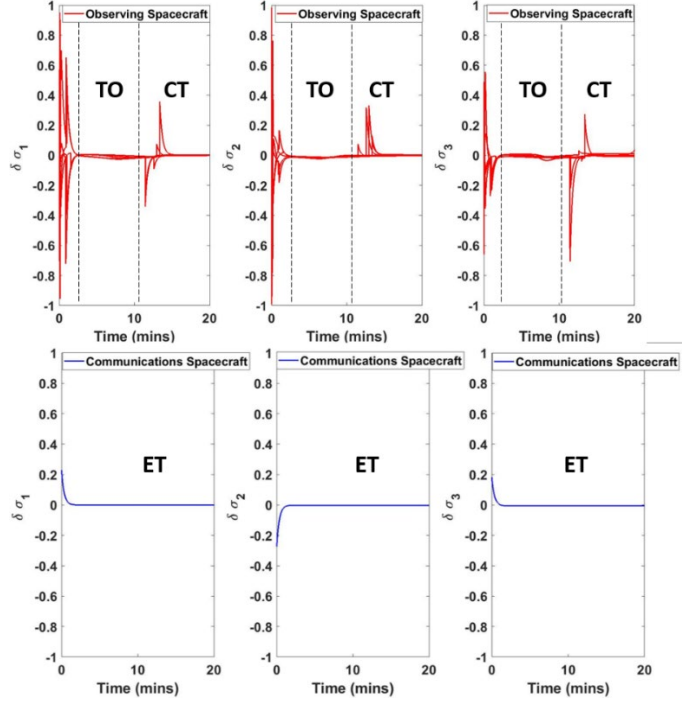

**Figure 9.** MRP Tracking errors of the swarm in Configuration 1. The tracking errors are shown to asymptotically converge to 0 indicating the reference is tracked.

*Configuration-2*. The MRP errors of the swarm in Configuration 2 is shown in Figure 10. In this simulation, Spacecraft 1 was the selected spacecraft which changes its behavior into the Communications spacecraft when in the transmitting region. As expected, the spike in the MRP error profiles of Spacecraft 1 indicates its switch from the target observing mode into an Earth tracking mode. The behavior switch is initiated as the spacecraft cross the imaging region and enter into the transmitting region. The controller is shown to regulate tracking errors. The remaining spacecraft continue their roles as the Observation Spacecraft. Their behaviors change from the observing the target surface to tracking the Communications Spacecraft as they exit the imaging region.

**Surface feature observation**

The surface observation of the 2 swarm configurations is visualized in Figure 11 by sampling the operation at $t = 10$ mins and $t = 20$ mins from their deployment. The direction of the spacecraft $z$ axis is denoted by dashed lines. At $t = 10$ mins all the spacecraft are inside the imaging region, while at $t = 20$ mins the spacecraft are completely in the transmitting region. When inside the imaging region, the Observation Spacecraft in Configuration 1 (colored red), are shown to observe the surface feature.

Once they enter the transmitting region, their $z$ axis is pointed towards the Communications Spacecraft. As expected, the Communications Spacecraft (colored blue) is always shown to track the LoS with respect to Earth. In Configuration 2, on the other hand, the selected Spacecraft 1 (colored blue) is shown to observe the surface feature when it is inside the imaging region.

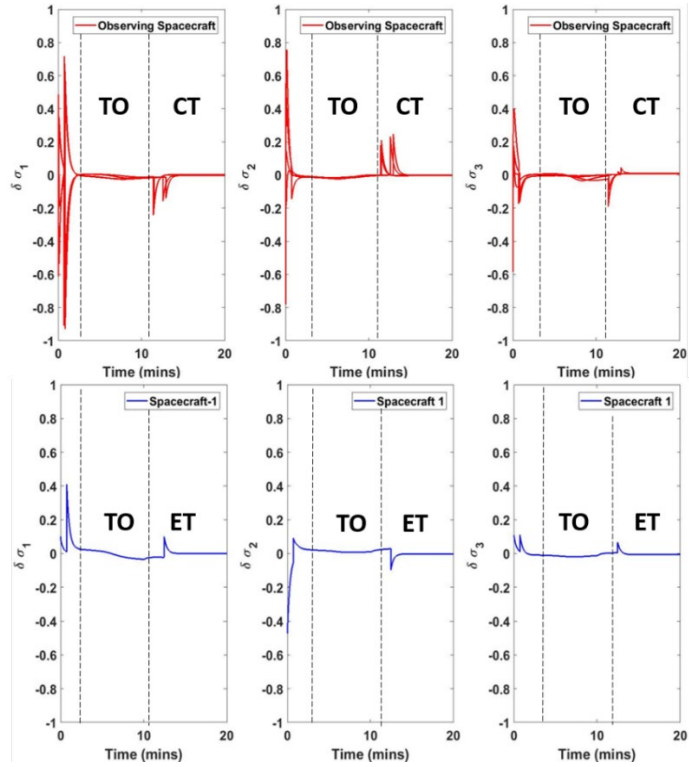

**Figure 10.** MRP Tracking errors of the swarm in Configuration 2. The tracking errors are shown to asymptotically converge to 0 indicating the reference is tracked.

As expected, Spacecraft 1 switches to point towards earth it enters the transmission region. The remaining spacecraft in the swarm (colored red) are now shown to follow the attitude behaviors of the Observation Spacecraft in Configuration 1. The target feature (Figure 6) is completely colored blue in both configurations at $t = 20$ mins, indicating 100 % coverage of the target surface during the flybys.

**DISCUSSION**

The simulations shown highlight a few important capabilities of the evolved swarm which will be discussed here. Firstly, if the spatial distribution of the target region of interest is known, an optimal spacecraft swarm can be designed to provide detailed surveillance of the target region. However, if the distribution is unknown, the swarm can be configured to operate using a stochastic scheme to explore the unknown regions. Secondly, as observed here, the spacecraft are shown to utilize a decentralized control law to track their references. While the spacecraft did interact with each other during communications, the behavioral change occurred due to the flag raised by the imaging distance, which is a parameter specific to the individual spacecraft.

This feature highlights the concept of a Class 2 swarm, where each spacecraft in the swarm adapts its behavior. Additionally, in Configuration 2, Spacecraft 1 was chosen as the Communications Spacecraft without any loss of generality. Such a decision can be made dynamically through a clustering scheme. In addition to this, the current work demonstrated the utility of the Automated Swarm Designer module of the IDEAS software by converting a surface exploration problem to a design optimization problem which was solved by an Evolutionary Algorithm. Through this work, the LoS tracking behavior is added to the knowledge base of swarm behaviors.

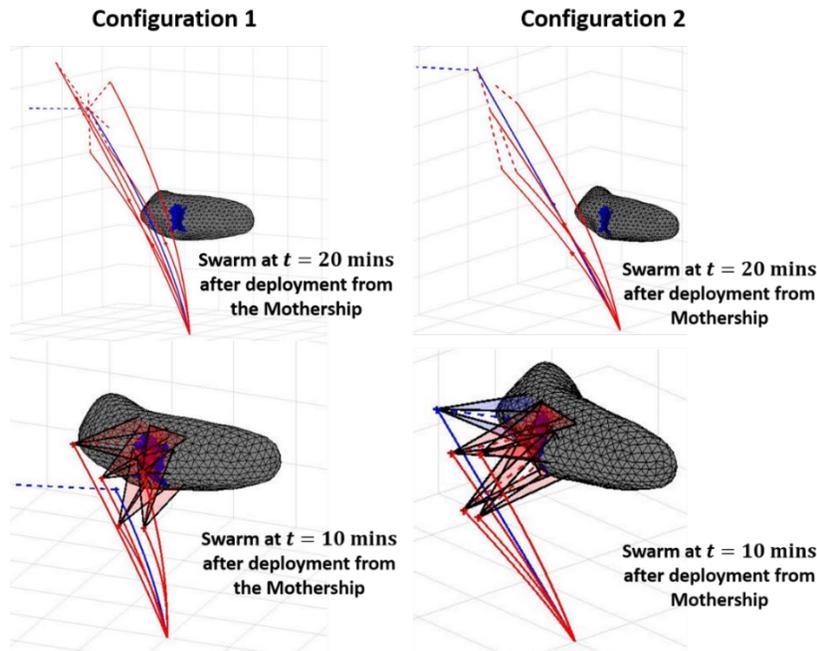

**Figure 11.** Visualization of the observation operation sampled at $t = 10$ mins and $t = 20$ mins after deployment from the mothership.

**CONCLUSIONS**

In this paper, we showed the utility of the Automated Swarm Design module of the IDEAS software to design a swarm which can observe a target feature of interest. The principles presented were then demonstrated through numerical simulations by designing a swarm which observes a hypothetical surface feature on the surface of asteroid 433 Eros. The optimal solution indicated that a swarm of 5 Observation Spacecraft would observe the target region completely. The operations of these spacecraft were simulated for the 2 configurations, and results showed successful performance of the control actions and automated swarm design principles. While the current work explored the design of a Class 2 swarm, future work will explore the designs of the remaining classes of swarms. Specifically, Classes 3 and 4 will require the design of formation flying and consensus-based controllers to regulate behavior. The inclusion of these tools will advance the capabilities of the IDEAS software towards designing efficient tools for exploring small bodies throughout the solar system.

**REFERENCES**

[1] Castillo-Rogez, J.C. et al., 2012. Expected science return of spatially-extended in-situ exploration at small Solar system bodies. *Aerospace Conference, 2012 IEEE*, pp.1–15

[2] Committee on the Planetary Science Decadal Survey, Space Studies Board & National Research Council, 2011. *Vision and Voyages for Planetary Science in the Decade 2013-2022*, National Academies Press.

[3] Nallapu, RT, Thoesen, A, Garvie, L, Asphaug, E & Thangavelautham, J 2016, 'Optimized bucket wheel design for asteroid excavation', *Proceedings of the International Astronautical Congress, IAC*

[4] Mazanek, D.D., Merrill, R.G., Brophy, J.R. and Mueller, R.P., 2015. Asteroid redirect mission concept: a bold approach for utilizing space resources. *Acta Astronautica*, 117, pp.163-171.

[5] Kalita, H., Ravindran, A. and Thangavelautham, J., 2018. Exploration and Utilization of Asteroids as Interplanetary Communication Relays. In *2018 IEEE Aerospace Conference*.

[6] Lantoine, G., Braun, R., Lantoine, G. and Braun, R.D., 2006. Optimal trajectories for soft landing on asteroids.

[7] Melman, J.C.P., Mooij, E. and Noomen, R., 2013. State propagation in an uncertain asteroid gravity field. *Acta Astronautica, 91*, pp.8-19.

[8] Nallapu, and Thangavelautham, J., 2019. Attitude Control of Spacecraft Swarms for Visual Mapping of Planetary Bodies. *2019 IEEE Aerospace Conference, 2019.*

[9] Coello, C.A.C., Lamont, G.B. and Van Veldhuizen, D.A., 2007. *Evolutionary algorithms for solving multi-objective problems* (Vol. 5). New York: Springer.

[10] Bandyopadhyay, S., Foust, R., Subramanian, G.P., Chung, S.J. and Hadaegh, F.Y., 2016. Review of formation flying and constellation missions using nanosatellites. *Journal of Spacecraft and Rockets, (0)*, pp.567-578.

[11] Wang, F., Nabil, A. and Tsourdos, A., 2009, May. Centralized/decentralized control for spacecraft formation flying near sun-earth l2 point. In. *4th IEEE Conference on* (pp. 1159-1166). IEEE.

[12] Ren, W. and Beard, R., 2004. Decentralized scheme for spacecraft formation flying via the virtual structure approach. *Journal of Guidance, Control, and Dynamics*, 27(1), pp.73-82.

[13] Wertz, J.R., 2001. Mission geometry *El Segundo, CA; Microcosm: Kluwer, 2001. Space technology library; 13.*

[14] Morgan, D., Chung, S.J., Blackmore, et al. 2012. Swarm-keeping strategies for spacecraft under J2 and atmospheric drag perturbations. *Journal of Guidance, Control, and Dynamics, 35*(5), pp.1492-1506.

[15] Alfriend, K., Vadali, S.R., Gurfil, et al. 2009. *Spacecraft formation flying* (Vol. 2). Elsevier.

[16] Agasid, E., et al. Small spacecraft technology state of the art. *NASA, Ames Research Center, Mission Design Division Rept. NASA/TP-2015-216648/REV1, Moffett Field, CA.*

[17] Morgan, D. et al., 2015. Guidance and control of swarms of spacecraft, Univ. of Illinois, Thesis.

[18] Lee, D., Sanyal, A.K. and Butcher, E.A., 2014. Asymptotic tracking control for spacecraft formation flying with decentralized collision avoidance. *Journal of Guidance, Control, and Dynamics, 38*(4), pp.587-600.

[19] Cornara, S., Beech, T.W., Belló-Mora, M. and Janin, G., 2001. Satellite constellation mission analysis and design. *Acta Astronautica, 48*(5-12), pp.681-691.

[20] Nallapu, R., Kalita, H. and Thangavelautham, J., 2018. On-Orbit Meteor Impact Monitoring Using CubeSat Swarms.

[21] Ely, T.A., Crossley, W.A. and Williams, E.A., 1999. Satellite constellation design for zonal coverage using genetic algorithms. *Journal of the Astronautical Sciences, 47*(3-4), pp.207-228.

[22] Curtis, S.A., Rilee, M.L., Clark, et al. 2003, Use of swarm intelligence in spacecraft constellations. In *Proceedings of the Third International Workshop on Satellite Constellations and Formation Flying* (pp. 24-26).

[23] Clark, P.E., Curtis, S., Rilee, M., Truszkowski, W. and Marr, G., 2002, March. ANTS: exploring the solar system with an autonomous nanotechnology swarm. In *Lunar and Planetary Science Conference* (Vol. 33).

[24] Vance, L., Asphaug, E., Thangavelautham, J., "Evaluation of Mother-Daughter Architectures for Asteroid Belt Exploration," *AIAA Science and Technology Forum*, pp. 1-8, 2019.

[25] Atchison, J., Mitch, R., Rivkin, A., NIAC Swarm Flyby Gravimetry Phase I Report.

[26] Mitch, R., Apland, C., Kee, C., Rivkin, A., Mazarico, E., Harclerode, K., Wortman, J. and Candela, K., NIAC Swarm Flyby Gravimetry Phase II Report.

[27] Scheeres, D.J., 2016. *Orbital motion in strongly perturbed environments*. Springer, Berlin, Germany.

[28] Misra, G., Izadi, M., Sanyal, A. and Scheeres, D., 2016. Coupled orbit–attitude dynamics and relative state estimation of spacecraft near small Solar System bodies. *Advances in Space Research*, 57(8), pp.1747-1761.

[29] Junkins, J.L. and Schaub, H., 2014. *Analytical mechanics of space systems*. AIAA, Washington DC.

[30] Kowalchuk, S. and Hall, C., 2008, August. Spacecraft attitude sliding mode controller using reaction wheels. In *AIAA/AAS Astrodynamics Specialist Conference and Exhibit* (p. 6260).

[31] Tsiotras, P., Shen, H. and Hall, C., 2001. Satellite attitude control and power tracking with energy/momentum wheels. *Journal of Guidance, Control, and Dynamics*, 24(1), pp.23-34.

[32] Conn, A.R., Gould, N.I. and Toint, P., 1991. A globally convergent augmented Lagrangian algorithm for optimization with general constraints and simple bounds. *SIAM Journal on Numerical Analysis*, 28(2), pp.545-572.

[33] Murray, C.D. and Dermott, S.F., 1999. *Solar system dynamics*. Cambridge university press.

[34] Heidt, H., Puig-Suari, J., Moore, A., Nakasuka, S. and Twiggs, R., 2000. CubeSat: A new generation of picosatellite for education and industry low-cost space experimentation.

[35] Pothamsetti, R., Thangavelautham, J., "Photovoltaic Electrolysis Propulsion System for Interplanetary CubeSats", Proceedings of the IEEE Aerospace Conference, 2016.

[36] Rabade, S., Barba, N., Garvie, L., Thangavelautham, J., "The Case for Solar Thermal Steam Propulsion System for Interplanetary Travel, " Proceedings of the 67th International Astronautical Congress, 2016.